\documentclass[aps,pra,twocolumn,showpacs,amsmath,amssymb,floatfix,footinbib,superscriptaddress]{revtex4-1}
\usepackage{graphics,graphicx,epsfig,ulem}
\usepackage{amsmath,amssymb,latexsym,float}
\usepackage{epstopdf,bm,longtable,url}
\usepackage{amsfonts,subfigure,hyperref,color}
\usepackage[ansinew]{inputenc}
\usepackage[usenames,dvipsnames]{pstricks}
\definecolor{darkblue}{rgb}{0.5,0.5,0.9}
\hypersetup{colorlinks,breaklinks,linkcolor=blue,urlcolor=blue,anchorcolor=blue,citecolor=blue}
\begin{document}
\baselineskip=14pt
\title{Large and tunable negative refractive index via electromagnetically induced chirality in a semiconductor quantum well nanostructure}

\author{Shun-Cai Zhao}
\email{zhaosc@kmust.edu.cn}
\affiliation{Physics department, Kunming University of Science and Technology, Kunming, 650500, PR China}

\author{Shuang-Ying Zhang}
\affiliation{Physics department, Kunming University of Science and Technology, Kunming, 650500, PR China}

\author{You-Yang Xu}%
\affiliation{Physics department, Kunming University of Science and Technology, Kunming, 650500, PR China}

\date{\today}

\begin{abstract}
Large and tunable negative refractive index (NRI) via electromagnetically induced chirality is
demonstrated in a semiconductor quantum wells (SQWs) nanostructure by using the reported experimental
parameters in Ref.[19]. It is found: the large and controllable NRI with alterable frequency
regions is obtained when the coupling laser field and the relative phase are modulated, which
will increase the flexibility and possibility of implementing NRI in the SQWs nanostructure. The scheme rooted in the
experimental results may lead a new avenue to NRI material in solid-state nanostructure.

\end{abstract}


\maketitle
\section{Introduction}
In recent years there has been a huge interest in semiconductor quantum wells(SQWs),
on account of the intersubband transitions (ISBTs) in SQWs are believed to have great
potential applications in solid-state optoelectronics and quantum information
science\cite{1,2,3,4}. In many case the characteristics of ISBT dephasing mechanisms
in SQWs behave essentially as ``artificial atoms'', which allowed us to regard the SQW
as a single quantum object for many kinds of quantum optical phenomena, in which the nonlinear
quantum optical phenomena have been extensively discussed, such as
Kerr nonlinearity\cite{5}, ultrafast all optical switching\cite{6}, coherent population
trapping\cite{7}, electromagnetically induced transparency\cite{8,9,10}, gain without
inversion\cite{11}, enhancing index of refraction\cite{12} and other novel phenomena
\cite{13,14,15,16,17,18}. But not only that, the devices based on ISBTs in SQWs
also have many inherent advantages that the atomic systems don't have, such as the high
nonlinear optical coefficients, the large transition energies and electric dipole moments
for the small effective electron mass, the great flexibilities in symmetries design as
well as in devices design by choosing the materials and structure dimensions.

\section{Theoretical Model}

In this work, we investigate the refractive index in a bulk SQW nanostructure via
electromagnetically induced chirality. And the results demonstrate the large and
tunable NRI can be realized by the coupling laser field and the relative phase.
The SQW samples we simulate here are reported to be grown by the molecular beam
epitaxy (MBE) method and each contains a 4.8nm $In_{0.47}Ga_{0.53}As$/ 0.2nm
$Al_{0.48}In_{0.52}As$/ 4.8nm$In_{0.47}Ga_{0.53}As$ coupled quantum well,
separated by a 36nm $Al_{0.48}In_{0.52}As$ modulation-doped barriers\cite{19,20}.
And the transition energies of the ISBTs were measured as $\omega_{12}$=124 meV, $\omega_{23}$=185 meV,
implying $\omega_{13}$=309 meV\cite{19}. The corresponding transition
dipoles were calculated as $d_{12}$=2.335 nm, $d_{23}$=2.341 nm, and
$d_{13}$ =0.120 nm\cite{21}. So the synthesized 3-level cascade electronic system of
ISBTs in such SQW forms a familiar ladder configuration, as shown in Fig.1.

\begin{figure}[htp]
\center
\includegraphics[totalheight=1.6 in]{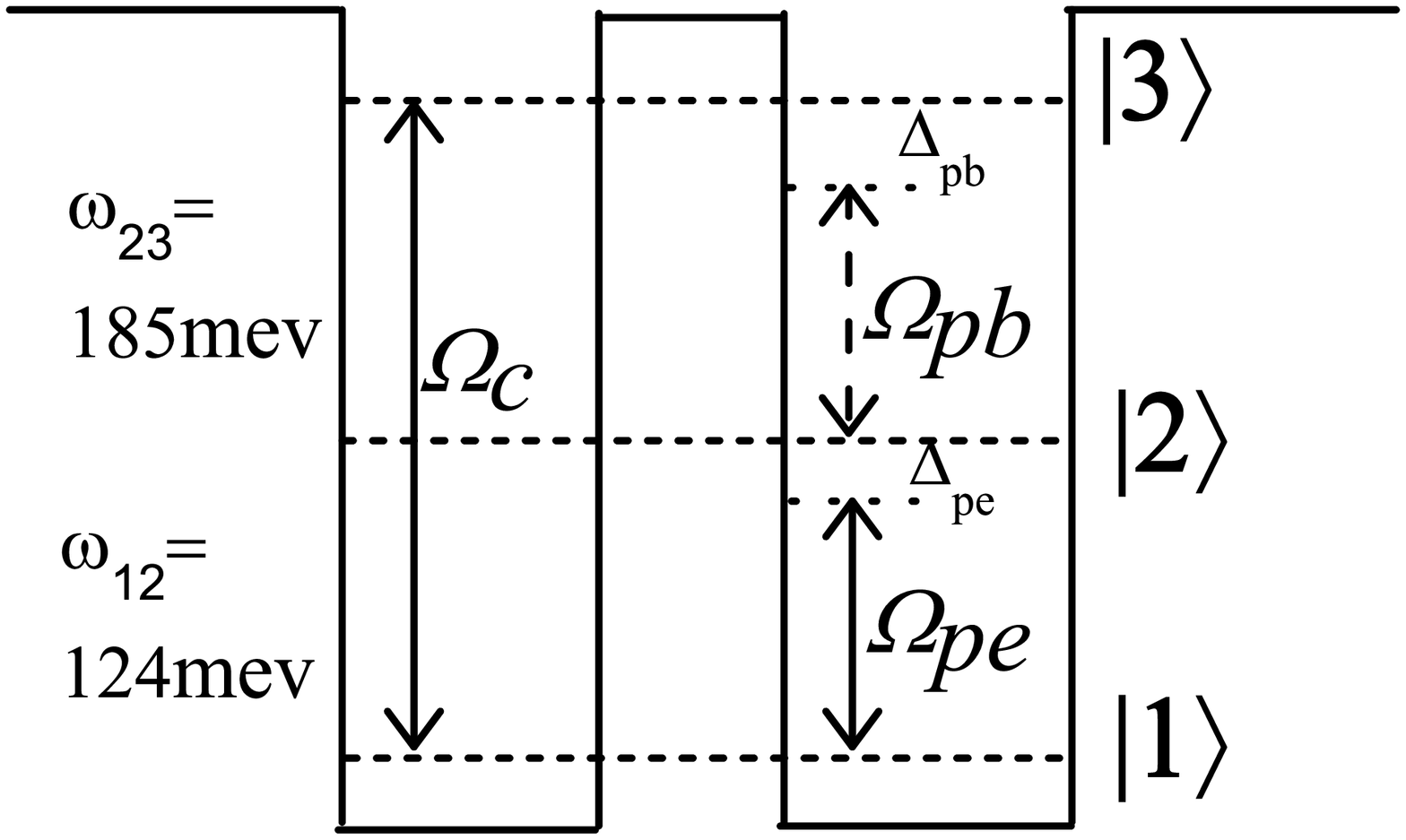 }
\caption{$\mathbf{Fig.1}$   Schematic diagram of the SQW structure. The electric-dipole transition
$|1\rangle$-$|3\rangle$ is driven by the  coupling laser field, and the level pairs
$|1\rangle$-$|2\rangle$ and $|3\rangle$-$|2\rangle$ are coupled
to the electric and magnetic fields of the weak probe light, respectively.}
\end{figure}\label{Fig.1}
The equivalent 3-level loop `` atom'' system simulated in our work is mainly
based on the above reported SQW sample. In such a 3-level system, the parity of level
$|1\rangle$ is set to be opposite to those of the levels $|2\rangle$ and $|3\rangle$, which
have the same parity. The possible optical transition $|1\rangle$-$|3\rangle$
is mediated by a coupling laser field with central frequency $\omega_{c}$ and
Rabi frequency $\Omega_{c}$. A weak probe laser field with central frequency $\omega_{p}$
and Rabi frequency $\Omega_{pe}$ is applied to the ISBT $|1\rangle$-$|2\rangle$.
Because of the parity selection rules, the two levels $|1\rangle$ and $|2\rangle$
with electric dipole element $d_{12}$ =$\langle2|$$\hat{\vec{d}}$$|1\rangle$$\neq0$ are
coupled by the electric component of the weak probe field, where $\hat{\vec{d}}$
is the electric dipole operator. The two levels $|2\rangle$ and $|3\rangle$ with the magnetic
dipole element $\mu_{23}$ =$\langle2|$$\hat{\vec{\mu}}$$|3\rangle$$\neq0$ are coupled
by the magnetic component of the probe field with Larmor frequency $\Omega_{pb}$=$\vec{B}_{p}\mu_{23}/\hbar$,
where $\hat{\vec{\mu}}$ is the magnetic-dipole operator. The coherent cross-coupling between
electric and magnetic dipole transitions driven by the electric and magnetic components
of the probe field may lead to chirality\cite{25,26}. Because the transition $|1\rangle$-$|3\rangle$
is mediated by a coupling laser field, the transitions $|1\rangle$-$|2\rangle$ and
$|3\rangle$-$|2\rangle$ are coupled respectively by the electric and magnetic components
of the probe field; it then forms a closed-loop system. It is well known that in a loop
configuration the dynamics behaviour becomes quite sensitive to phases\cite{22}.

Under the rotating-wave approximations in the interaction representation, with the
assumption of $\hbar$=1 and employing the quantum regression theorem and the
established treatment for 3-level atoms\cite{23}, the resulting Hamiltonian of
the three-level configuration can be written as
\begin{eqnarray}
&H_{int} =&\Delta_{pe}|2\rangle\langle2|+(\Delta_{c}+\Delta_{pb})|3\rangle\langle3|-(\Omega_{pe}exp(-i\varphi_{pe}) \nonumber\\
          &&|2\rangle\langle1|+\Omega_{pb}exp(-i\varphi_{pb})|3\rangle\langle2|+\Omega_{c} \label{eq1}\\
          &&exp(-i\varphi_{c})|3\rangle\langle1|+H.c), \nonumber
\end{eqnarray}
where $\Delta_{pe}$=$\omega_{21}$-$\omega_{p}$, $\Delta_{pb}$=$\omega_{23}$-$\omega_{p}$
and $\Delta_{c}$=$\omega_{13}$-$\omega_{c}$ are the ISBTs detunings of the corresponding fields, and
$\Delta_{c}$ depicts the two photon detuning process. According to the measured transition
energies in the above SQW sample\cite{19}, we can get the relationship $\Delta_{c}$=$\Delta_{pe}$ + $\Delta_{pb}$,
which can avert the major obstacle mentioned in Ref.\cite{24} in realizing the predicted effects at a realistic
experimental setting. The $\varphi_{pb}$, $\varphi_{pe}$ and $\varphi_{c}$ are denoted the relevant phases of the three
coherent fields. The symbol H.c. means the Hermitian conjugate. Then the equation of
the time-evolution for the system can be described as  $\frac{d\rho}{dt}=-\frac{i}{\hbar}[H,\rho]+\Lambda\rho $,
where $\Lambda\rho$ represents the irreversible decay part in
the system. Under the dipole approximation the density matrix equations described the system are
written as follows:
\begin{eqnarray}
&\dot{\rho_{22}}=&i\Omega_{pb}(\rho_{32}-\rho_{23})+i\Omega_{pe}(\rho_{12}-\rho_{21})-\gamma_{21}\rho_{22} \nonumber\\
                 &&+\gamma_{32}\rho_{33},\nonumber\\
&\dot{\rho_{33}}=&i\Omega_{pb}(\rho_{32}-\rho_{23})+i\Omega_{c}exp(-i\phi)\rho_{13}-i\Omega_{c} \nonumber\\
                 &&exp(i\phi)\rho_{31}-(\gamma_{32}+\gamma_{31})\rho_{33},\nonumber\\
&\dot{\rho_{12}}=&i\Omega_{pe}(\rho_{22}-\rho_{11})+i\Omega_{c}exp(i\phi)\rho_{32}-i\Omega_{pb} \nonumber\\
                 &&(\rho_{13}+(i\Delta_{pe}-\frac{\Gamma_{12}}{2})\rho_{12}, \label{eq2}\\
&\dot{\rho_{13}}=&i\Omega_{c}exp(i\phi)(\rho_{33}-\rho_{11})+i\Omega_{pe}(\rho_{23}-i\Omega_{pb}  \nonumber\\
                 &&\rho_{12}+[i(\Delta_{pe}+\Delta_{pb})-\frac{\Gamma_{13}}{2}]\rho_{13},\nonumber\\
&\dot{\rho_{23}}=&i\Omega_{pb}(\rho_{33}-\rho_{22})+i\Omega_{pe}\rho_{13}-i\Omega_{c}(\rho_{21}  \nonumber\\
                 &&+(i\Delta_{pb}-\frac{\Gamma_{23}}{2})\rho_{23},\nonumber
\end{eqnarray}
where $\phi$=$\varphi_{c}-\varphi_{pe}-\varphi_{pb}$ is the relative phase of the three corresponding
optical fields. And the above density matrix elements comply with the conditions: $\rho_{11}+\rho_{22}+\rho_{33}$=$1 $
and $\rho_{ij}$=$\rho_{ji}^{\ast}$. The total decay rates $\Gamma_{ij}$ are added
phenomenologically\cite{19,25} in above density matrix equations, which include the
population decay rates and dephasing decay rates. Among them, the population decay rates
from ISBTs, denoted by $\gamma_{ij}$, are primarily due to
longitudinal optical(LO) phonon emission events at low temperature.
And the dephasing decay rates originate from electron-electron, interface roughness
and phonon scattering processes. Thus the total decay rates $\Gamma_{ij}$ can been written by
 $\Gamma_{12}$=$\gamma_{12}+\gamma^{dph}_{12}$, $\Gamma_{23}$=$\gamma_{31}+\gamma_{32}+\gamma_{21}+\gamma^{dph}_{23}$,
 $\Gamma_{13}$=$\gamma_{31}+\gamma_{32}+\gamma^{dph}_{13}$. When the probe field is weak,
 i.e. $\Omega_{pe}$, $\Omega_{pb}$$\ll$$\Omega_{c}$, $\Gamma_{ij}$, almost all the atoms can be
 assumed to be in the ground state$|1\rangle$, the steady-state values of the density matrix
 elements $\rho_{12}$ and $\rho_{23}$ can be written in a linear approximation
\begin{equation}
\rho_{12}=\frac{2A_{2}d_{12}}{iA_{1}A_{3}}\textbf{E}+\frac{4e^{-i\phi}\Omega_{c}(\Gamma_{23}+2i\Delta_{pb})u_{23}}{A_{1}A_{3}}\textbf{B},
\end{equation}
\begin{equation}
\rho_{23}=\frac{4e^{i\phi}A_{4}\Omega_{c}d_{12}}{A_{2}A^{*}_{3}}\textbf{E}+\frac{8u_{23}e^{2i\phi}\Omega^{2}_{c}}{iA^{*}_{1}A^{*}_{3}}\textbf{B}
\end{equation}
where $A^{*}_{i}(i$=$1,2,3,4)$ is the conjugate complex of $A_{i}$, and with $A_{1}$=$\Gamma_{12}(\Gamma_{23}+2i\Delta_{pb})-2i\Gamma_{23}\Delta_{pe}+4(\Delta_{pe}\Delta_{pb}+\Omega^{2}_{c}),$
$A_{2}$=$\Gamma_{13}(\Gamma_{23}+2i\Delta_{pb})+\Gamma_{12}(\Gamma_{23}+2i\Delta_{pb})-A_{1}-4\Delta^{2}_{pb},$
$A_{3}$=$\Gamma_{13}+2i(\Delta_{pe}+\Delta_{pb})$, $A_{4}$=$\Gamma_{12}+\Gamma_{12}-2i\Delta_{pb}.$ The ensemble electric
polarization and magnetization of the bulk SQWs
to the probe field are given by $\vec{P}$=$N\vec{d_{12}}\rho_{21}$ and
$\vec{M}$=$N\vec{\mu_{23}}\rho_{32}$, respectively, where N is the
density of SQWs. Then the coherent cross-coupling between electric
and magnetic dipole transitions driven by the electric and magnetic
components of the probe field may lead to chirality\cite{26,27}.
Substituting equations (3) and (4) into the formula for the ensemble
electric polarization ($\vec{P}$=$N\vec{d_{12}}\rho_{21}$) and
magnetization ($\vec{M}$=$N\vec{\mu_{23}}\rho_{32}$), we have the relations
$\vec{P}$=$\alpha_{EE} \vec{E}+\alpha_{EB} \vec{B}$,
$\vec{M}$=$\alpha_{BE} \vec{E}+\alpha_{BB} \vec{B}$.
Considering both electric and magnetic local field effects\cite{28}, $\vec{E}$ and $\vec{B}$ in
$\vec{P}$ and $\vec{M}$ must be replaced by the local fields:
$\vec{E}_{L}$=$\vec{E}+\frac{\vec{P}}{3\varepsilon_{0}}, \vec{B}_{L}$=$\mu_{0}(\vec{H}+\frac{\vec{M}}{3})$.
Then we obtain
\begin{eqnarray}
&\vec{P}=&\frac{3\varepsilon_{0}(\mu_{0}\alpha_{BB}\alpha_{EE}-\mu_{0}\alpha_{BE}\alpha_{EB}-3\alpha_{EE})}{B_{1}}\vec{E}\nonumber\\
            &&+\frac{-9\mu_{0}\varepsilon_{0}\alpha_{EB}}{B_{1}}\vec{H},\\
&\vec{M}=&\frac{3(\mu_{0}\alpha_{BE}\alpha_{EB}-\mu_{0}\alpha_{BB}\alpha_{EE}+3\varepsilon_{0}\mu_{0}\alpha_{BB})}{B_{2}}\vec{H}\nonumber\\
            &&+\frac{9\varepsilon_{0}\alpha_{BE}}{B_{2}}\vec{E} \nonumber
\end{eqnarray}
where $B_{1}$=$\mu_{0}\alpha_{BE}\alpha_{EB}+3\alpha_{EE}-\mu_{0}\alpha_{BB}\alpha_{EE}-9\varepsilon_{0}+3\mu_{0}\varepsilon_{0}\alpha_{BB}$,
$B_{2}$=$\mu_{0}\alpha_{BB}\alpha_{EE}+9\varepsilon_{0}-\mu_{0}\alpha_{BE}\alpha_{EB}-3\alpha_{EE}-3\varepsilon_{0}\mu_{0}\alpha_{BB}$ with
$\alpha_{EE}$=$-2[\Gamma_{13}(\Gamma_{23}+2i\Delta_{pb})+2i\Gamma_{23}(\Delta_{pb}+\Delta_{pe})-4(\Delta^{2}_{pb}+\Delta_{pb}\Delta_{pe}
+\Omega^{2}_{c})]d_{12}/[2(\Delta_{pe}+\Delta_{pb})-i\Gamma_{13}][\Gamma_{12}(\Gamma_{23}+2i\Delta_{pb})-2i\Gamma_{23}\Delta_{pe}+4(\Delta_{pe}\Delta_{pb}+\Omega^{2}_{c})]\hbar$,
$\alpha_{EB}$=$4e^{-i\phi}(\Gamma_{23}+2i\Delta_{pb})\Omega_{c}\mu_{23}/[\Gamma_{13}+2i(\Delta_{pe}+\Delta_{pb})][\Gamma_{12}(\Gamma_{23}+2i\Delta_{pb})-2i\Gamma_{23}\Delta_{pe}+4(\Delta_{pe}\Delta_{pb}+\Omega^{2}_{c})]\hbar$,
$\alpha_{BE}$=$4e^{i\phi}(\Gamma_{12}+\Gamma_{13}-2i\Delta_{pb})\Omega_{c}d_{12}/[\Gamma_{13}-2i(\Delta_{pe}+\Delta_{pb})][\Gamma_{12}(\Gamma_{23}-2i\Delta_{pb})+2i\Gamma_{23}\Delta_{pe}+4(\Delta_{pe}\Delta_{pb}+\Omega^{2}_{c})]\hbar$,
$\alpha_{BB}$=$8e^{2i\phi}\Omega^{2}_{c}\mu_{23}/[\Gamma_{13}-2i(\Delta_{pe}+\Delta_{pb})][\Gamma_{12}(i\Gamma_{23}+2\Delta_{pb})-2\Gamma_{23}\Delta_{pe}+4i(\Delta_{pe}\Delta_{pb}+\Omega^{2}_{c})]\hbar$.
The key idea of electromagnetic induced chirality is to
use the magnetoelectric cross-coupling in which the electric polarization
$\vec{P}$ is coupled to the magnetic field $\vec{H}$ of an
electromagnetic wave and the magnetization $\vec{M}$ is coupled to
the electric field $\vec{E}$ \cite{27}:
\begin{equation}
\vec{P}=\varepsilon_{0}\chi_{e} \vec{E}+\frac{\xi_{EH}}{c} \vec{H},\\
\vec{M}=\frac{\xi_{HE}}{c \mu_{0}} \vec{E}+\chi_{m} \vec{H}
\end{equation}
Here $\chi_{e}$ and $\chi_{m}$, $\xi_{EH}$ and $\xi_{HE}$ are
the electric and magnetic susceptibilities, and the complex chirality
coefficients, respectively. They lead to additional contributions to
the refractive index for one circular polarization\cite{27,29}:
\begin{equation}
n=\sqrt{\varepsilon\mu-\frac{(\xi_{EH}+\xi_{HE})^{2}}{4}}+\frac{i}{2}(\xi_{EH}-\xi_{HE})
\end{equation}

By comparison with equations (5) and (6), we obtain the permittivity by $\varepsilon$=$1+\chi_{e}$ and the
permeability by $\mu$=$1+\chi_{m}$, and the complex chirality coefficients $\xi_{EH}$ and $\xi_{HE}$.
In the above, we obtained the expressions for the electric permittivity and magnetic permeability
of the bulk SQWs. Substituting equations from (3) to (6) into (7), the expression
for refractive index can also be presented. In the section that follows, we will discuss
the refractive index of the SQWs with the experimental parameters from Ref.[19].

\section{Results and discussion}
In this work, some key parameters are selected from the Ref.\cite{19}, which
has the advantage than the photonic-resonant materials because of the
parameters used in the numerical simulation coming from the reported experimental results.
The key parameters of the above mentioned SQWs sample are the measured
transition energies\cite{19}, according to which the detunings of electric-dipole
and magnetic-dipole transitions are defined by the relation
$\Delta_{pb}$=$\Delta_{pe}$+$61$ mev, so we depict the two transitions to be different by
setting $\Delta_{pb}$$\neq$$\Delta_{pe}$ (i.e. two transition frequencies
are not near the same frequency in this SQWs  system)\cite{24}.
In order to ensure linear response, the probe laser field is kept more
than 100 times weaker to the coupling laser field.  As far as the mentioned bulk
SQWs sample be concerned, we estimate its average density in the cubic volume element
as N$\approx$$1.04\times10^{22}$$m^{-3}$ from the mentioned experimental parameters.
The parameter for the electric transition dipole moment from $|2\rangle$ $\leftrightarrow$ $|1\rangle$
is chosen from the measured parameter: $ d_{12}=2.335\times1.602\times10^{-19}C m$, and the magnetic transition
dipole moment is chosen from the typical parameter $ \mu_{23}=7.0\times10^{-23}Cm^{2}s^{-1}$\cite{30}.
Thus the total decay rates $\Gamma_{ij}$ are set $\Gamma_{ij}$=5 mev for all three ISBTs
transitions from the measured parameters\cite{19}. Another parameter is the coupling laser
intensity, which should be below the damage threshold of SQWs. For its
Rabi frequency, in this paper we choose the ranges of $\Omega_{c}$$\leq$ 20 meV.
For example, when taking the maximal Rabi frequency $\Omega_{c}$= 20 meV to make a calculation, the electric
field amplitude is obtained $E\approx10^{5}V/cm$ according to the relationship $\hbar\Omega=\mu E$.
Using of the connection between the electric field amplitude E and the intensity of the radiation I :
E=27.4682$\times\sqrt{I}$, we can get the intensity of the radiation I=13 MW/$cm^{2}$, which shows
the above-mentioned laser intensities may be satisfied below the damage threshold of quantum object\cite{31}.
For the sake of simplification, we take the unit with $\varepsilon_{0}$ = $\mu_{0}$ = 1.

Firstly, we discuss the refractive index dependence on the coupling laser
field's Rabi frequences. We concentrate on the situation when the
negative refraction is most prominent, i.e., under the condition of $\phi$ = 0.
In Fig. 2, the plots of Re[n] is shown as a function of $\Delta_{pe}$ with different $\Omega_{c}$=
9 mev, 11 mev, 13 mev and 15 mev. And the dotted line, dashed dotted line, dashed line and solid
line correspond to the four different values of $\Omega_{c}$, respectively. The dotted line has
the most narrow interval of [-62.7mev, 58.3mev], in which the refraction index is negative
and its maxima is -0.7. The dashed dotted line shows its maxima of -1.32 in the interval
[-64.5 mev, -56.65 mev] for NRI. The maxima of NRI arrives to -2.3 in the interval [-64.0 mev, -54.5 mev]
when $\Omega_{c}$ was tuned to 13 mev. The interval expands to [-64.9 mev, -51.8 mev]
in which the maxima of NRI is -3.7 when $\Omega_{c}$=15 mev. The increasing maxima of NRI in these
expanding intervals is obtained by the gradually increasing $\Omega_{c}$, which can be explained via
the quantum interference and coherence. The increasing coupling laser field drives the transition
$|1\rangle$-$|3\rangle$ in the SQWs, which enhances interference between the electric dipole element
and magnetic dipole element from the probe laser in ISBTs, then modifies the refractive index properties of the SQWs sample.
The adjustable bandwidth for NRI in different frequency regions can also be drawn from equations (3) and (4).
In equations (3), the coherent term $\rho_{12}$ is composed of two items: the former item is independent
$\Omega_{c}$, while the latter cross-coupling item is proportional to $\Omega_{c}$ then be sensitive to
$\Omega_{c}$. In the coherent term $\rho_{23}$, the two parts are proportional to
$\Omega_{c}$ and $\Omega^{2}_{c}$, respectively. Then the induced chirality depends strongly
on $\Omega_{c}$, which causes the refractive index being negative and the variable bandwidths.

\begin{figure}[htp]
\center
\includegraphics[totalheight=1.6 in]{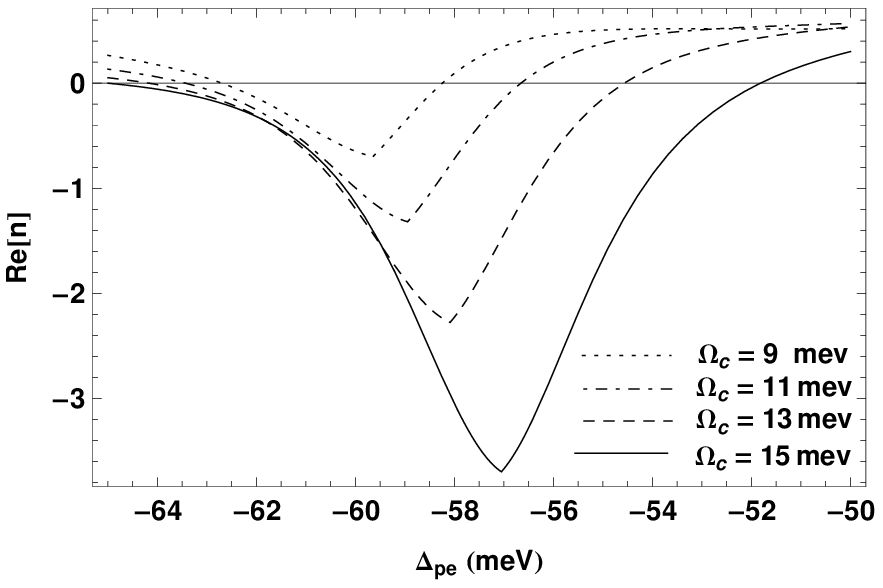 }
\caption{$\mathbf{Fig.2}$   The real parts of the refractive index n as a function of the probe detuning
$\Delta_{pe}$ with different Rabi frequencies $\Omega_{c}$ of coupling laser field:
9 mev(dotted line), 11 mev(dashed dotted line), 13 mev(dashed line), 15 mev(solid line).
The parameters are: $\phi$=0, $\Gamma_{12}$=$\Gamma_{13}$=$\Gamma_{23}$=5emv,
$\Omega_{pe}$=0.02emv, $\Delta_{pb}$=$\Delta_{pe}$+61 mev.}
\end{figure}\label{Fig.2}

After studying the refractive index dependence on the coupling laser field's
Rabi frequencies, we next study how the relative phase $\phi$ bring changes
in the refractive index of the SQWs sample.
As mentioned before, the dynamics behaviour is quite sensitive to phases\cite{22} in
a loop configuration, as can be see from equations (3) and (4). Equation (3) shows
that the former coefficient in the coherent term $\rho_{12}$ is independent of the
relative phase $\phi$, while the latter coefficient is sensitive to the relative phase $\phi$ through $e^{-i \phi}$.
Similarly, equation (4) has the phase-dependent chirality expression before coefficient $\vec{E}$
through $e^{ i \phi}$ and its second chirality part is dependent
of the relative phase $\phi$ with $e^{2i \phi}$. Thus, due to the presence
of the induced chirality depends strongly on the relative phase $\phi$, which causes
the refractive index to be negative via the adjusting relative phase.

\begin{figure}[htp]
\center
\includegraphics[totalheight=1.6 in]{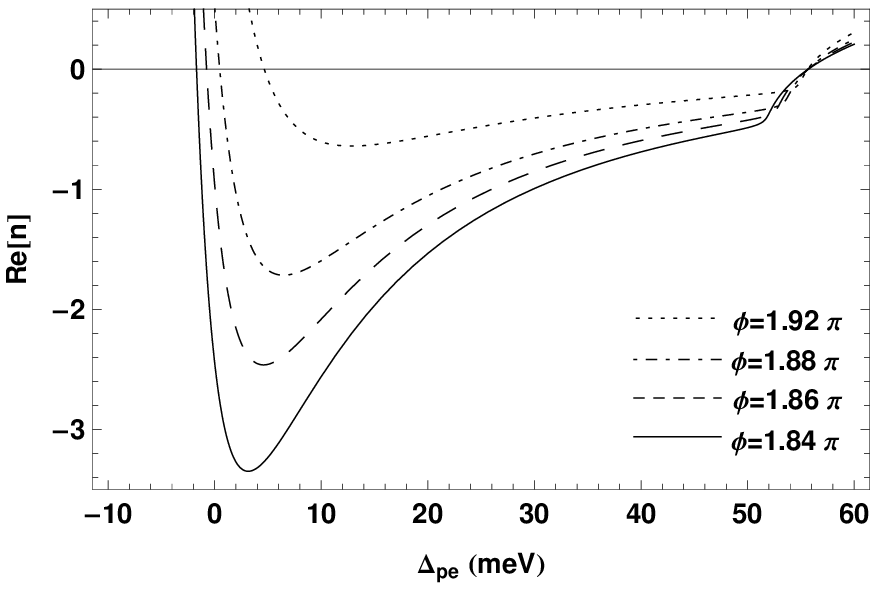 }
\caption{$\mathbf{Fig.3}$   The real parts of the refractive index n as a function of the probe
detuning $\Delta_{pe}$ with different values of relative phase $\phi$: 1.92$\pi$(dotted line),
1.88$\pi$(dashed dotted line), 1.86$\pi$(dashed line) and 1.84$\pi$(solid line),
$\Omega_{c}$=20mev. The other parameters are the same as Fig.2.}
\end{figure}\label{Fig.3}

In Fig. 3, the refraction index Re[n] is plotted as a function of $\Delta_{pe}$ with the relative phase
$\phi$ being 1.92$\pi$(dotted line), 1.88$\pi$(dashed dotted line), 1.86$\pi$(dashed line) and
1.84$\pi$(solid line) and $\Omega_{c}$=20mev. The sensitive adjustability via the relative phase
$\phi$ is shown by the curves in Fig. 3. Interestingly, the Re[n] varies significantly with the different values of $\phi$
i.e.,  $\phi$=1.92$\pi$, 1.88$\pi$, 1.86$\pi$ to 1.84$\pi$ when $\Omega_{c}$=20mev. The amplitudes of variation in the relative
phase $\phi$ are 0.04$\pi$, 0.02$\pi$ and 0.02$\pi$ between the dotted line,
dashed dotted line, dashed line and solid line, which causes the maximum of NRI changing from -0.65, -1.78, -2.50 to -3.45, and the corresponding intervals enlarge from [4.6 mev, 55.68 mev], [0.48 mev, 55.68 mev], [-0.78 mev, 55.68 mev] to [-1.70 mev, 55.68 mev]. Hence, the bandwidths for NRI can also be adjusted by the relative phase. Comparing Fig.2 with Fig.3,
the relative phase is much more sensitive than the coupling laser field in controlling the bandwidths for NRI.

\section{Conclusion}
In summary, the large and tunable NRI is theoretically demonstrated in a SQWs nanostructure
by using the reported experimental parameters. In our scheme, the coherent cross-coupling
between electric and magnetic transitions leads to chirality and the chirality coefficients
are modulated by the coupling laser field and relative phase, so the SQWs nanostructure can become
NRI material. When the coupling laser field and relative phase are properly modulated, the large and tunable NRI with
alterable frequency regions can be obtained. The flexibly modulating parameters and using reported
experimental parameters increase the possibility of implementing NRI in the SQWs nanostructure,
which may give us a newt way to NRI material in solid state nanotructure. And we hope the
coming experiment will achieve this in the future.

\section{Acknowledgment}
The work is supported by the National Natural Science Foundation of
China (Grant No. 61205205)and the Foundation for Personnel training
projects of Yunnan Province, China (Grant No. KKSY201207068).

\end{document}